\documentclass[twocolumn,aps]{revtex4-1}
\usepackage[latin9]{inputenc}
\usepackage{verbatim}
\usepackage{amsmath}
\usepackage{amssymb}
\usepackage{graphicx}

\makeatletter
\usepackage{eqnarray}\usepackage{amsthm}
\usepackage{fancyhdr}
\usepackage{subfigure}\usepackage{epsfig}
\usepackage{epsfig}
\usepackage{bbm}
\usepackage{subfigure}
\usepackage{bm}
\usepackage{float}

\newcommand{\F}{\boldsymbol{F}}

\newcommand{\pp}{\boldsymbol{p}}

\newcommand{\p}{\boldsymbol{p}}

\newcommand{\rr}{\boldsymbol{r}}
\newcommand{\RR}{\boldsymbol{R}}

\newcommand{\xxi}{\boldsymbol{\xi}}

\newcommand{\cchi}{\boldsymbol{\chi}}
\newcommand{\eeta}{\boldsymbol{\eta}}
\newcommand{\R}{\boldsymbol{R}}

\newcommand{\f}{\boldsymbol{f}}

\makeatother

\begin{document}
\title{Chemotaxis of cargo-carrying self-propelled particles}
\author{Hidde D. Vuijk, Holger Merlitz, Michael Lang, Abhinav Sharma, Jens-Uwe
Sommer}
\begin{abstract}
Active particles with their characteristic feature of self-propulsion
are regarded as the simplest models for motility in living systems.
The accumulation of active particles in low activity regions has led
to the general belief that chemotaxis requires additional features
and at least a minimal ability to process information and to control
motion. We show that self-propelled particles display chemotaxis and
move into regions of higher activity, if the particles perform work
on passive objects, or cargo, to which they are bound. The origin
of this cooperative chemotaxis is the exploration of the activity
gradient by the active particle when bound to a load, resulting in
an average excess force on the load in the direction of higher activity.
Using a minimalistic theoretical model, we capture the most relevant
features of these active-passive dimers and in particular we predict
the crossover between anti-chemotactic and chemotactic behaviour.
Moreover we show that merely connecting active particles to chains
is sufficient to obtain the crossover from anti-chemotaxis to chemotaxis
with increasing chain length. Such an active complex is capable of
moving up a gradient of activity such as provided by a gradient of
fuel and to accumulate where the fuel concentration is at its maximum.
The observed transition is of significance to proto-forms of life
enabling them to locate a source of nutrients even in the absence
of any supporting sensomotoric apparatus.
\end{abstract}
\maketitle
Escherichia coli is able to steer toward sources of nutrients by altering
its tumble rate~\citep{berg_04}. While searching, it performs temporal
comparisons of nutrient concentrations along its trajectory, and as
the concentration increases, the swimmer lowers its tumble rate to
move up the concentration gradient of nutrients. This search strategy,
known as \textit{klinokinesis with adaption}, leads to accumulation
of bacteria near the top of the nutrient density and thus to chemotaxis~\citep{schnitzer_Symp90,berg_93}.
It requires a delicate apparatus for chemical sensing, information
processing and a corresponding motoric response, available only to
organisms that have reached an advanced level of evolution.

Much simpler are self-propelled particles, which merely adjust their
speed in response to the \emph{local} fuel concentration, a mechanism
called \textit{orthokinesis}. Synthetic Janus particles, an example
for this class of swimmers, are driven by catalytic reactions in the
presence of e.g.\ hydrogen peroxide~\citep{howse_PRL07} or hydrazine~\citep{gao_JACS14}.
Biological molecules such as catalytic enzymes have also been reported
to show orthokinetic behaviour~\citep{jee_PNAS18b,jee_PNAS18b}.
Instead of intentional tumble movements, induced by active cilia as
used by E.\ coli, these particles change direction randomly through
rotational Brownian motion, and, if of roughly spherical shape, the
rotational relaxation time is almost independent of the degree of
propulsion. Usually, their swim speed (i.e. their degree of activity)
is positively correlated with the concentration of fuel. As a consequence,
one can show that in presence of a fuel gradient such active Brownian
particles (ABPs) accumulate in regions where the fuel concentration
is low, known as anti-chemotactic behaviour~\citep{schnitzer_PRE93}.
This corresponds qualitatively to that of passive Brownian particles
in a temperature gradient \citep{kampen_ZFP87}, and virtually to
all everyday experience that physical bodies accumulate in regions
where they are less agitated or moved. This has been recently shown
to play a role in the positioning of the nucleus in certain animal
cells \citep{almonacid2015active,razin2017generalized,razin2017forces}.
Although ABPs can display interesting transient behaviour in fuel
gradients, called pseudo-chemotaxis~\citep{lapidus_JTheoBiol80,peng_AC15,ghosh_PRE15,vuijk_PRE18},
activity does not lead to an advantage in the search for fuel sources
as compared to passive diffusion on long time scales~\citep{merlitz_PlosOne20}.

\begin{figure}[t]
\includegraphics[width=0.95\columnwidth]{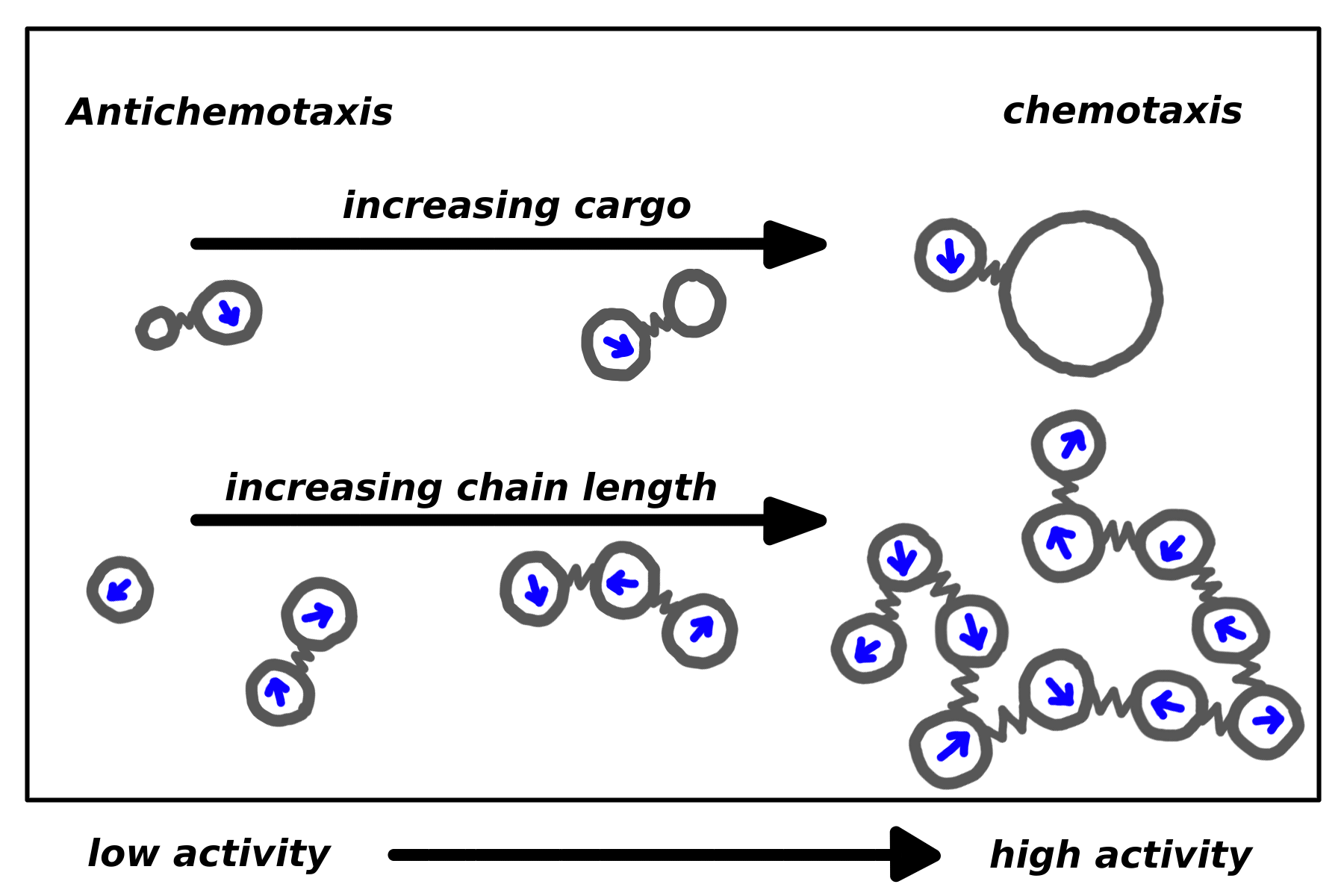} \caption{Crossover from anti-chemotactic to chemotactic behaviour. ABPs are
sketched as circles with arrows. Cargo particles (sketched as empty
circles) are bound to ABPs. (a) When the cargo is small, that is,
its friction coefficient is small, the dimer accumulates in low activity
regions (anti-chemotaxis). When the cargo is large, the dimer accumulates
in the high activity regions (chemotactis). (b) Self emergence of
cargo in active chains. When active particles are connected in a chain,
one observes the crossover from anti-chemotactic to chemotactic behaviour
with increasing chain length.}
\end{figure}

These observations have led to the general belief that chemotaxis,
a prominent feature of living systems, cannot be reflected by uninformed
moving objects, in particular not by ABPs. In this work, we demonstrate
that ABPs can show chemotactic behaviour in activity gradients when
they are bound to a passive cargo particle. With increasing cargo,
the active-passive complex switches its behaviour from anti-chemotactic
to chemotactic and accumulates in regions with a large fuel concentration
(see Fig. 1). Furthermore, we show that an explicit distinction between
cargo and active particles is not fundamental to our proposed mechanism
and can self-emerge in an active system. We demonstrate this in a
system of ABPs connected in a chain: with increasing chain length
there is a crossover from anti-chemotaxis to chemotaxis.

Our model consists of an ABP ~\citep{bechinger_RevModPhys16} attached
to a passive cargo by a stiff bond, thus forming a dimer. The activity
of the ABP is characterized by its persistence time $\tau$ (characteristic
time of rotation of the active particle) and a space-dependent swim
force $f_{s}(\rr)$, which we assume to be proportional to the local
fuel concentration. Regions where the swim force is large (small)
we call high (low) activity regions. The friction coefficient of the
ABP is $\gamma$ and that of the cargo is $q\gamma$. The load on
the ABP is increased by increasing $q$. Temperature is denoted by
$T$ in units such that the Boltzmann constant is unity. In our theoretical
approach, we coarse grain the Fokker-Planck equation for the dimer
to obtain an effective equation for the 'center-of-friction' coordinate
$\R$ defined as the friction-weighted vector sum of the position
vectors of the ABP and the load. In this letter, we consider the regime
where both the bond length as well as the persistence length of the
ABP are small compared to gradients of the swim force. Under these
assumptions, the distribution of the ABP is approximately the same
as that of the load and therefore $\R$ alone can be used to specify
the location of the dimer. The steady state distribution of the dimer
is obtained by setting the fluxes to zero in the effective equation
(see Methods). Our results hold for $d\in\{2,3\}$ dimensions.

\begin{figure}
\includegraphics[width=1\columnwidth]{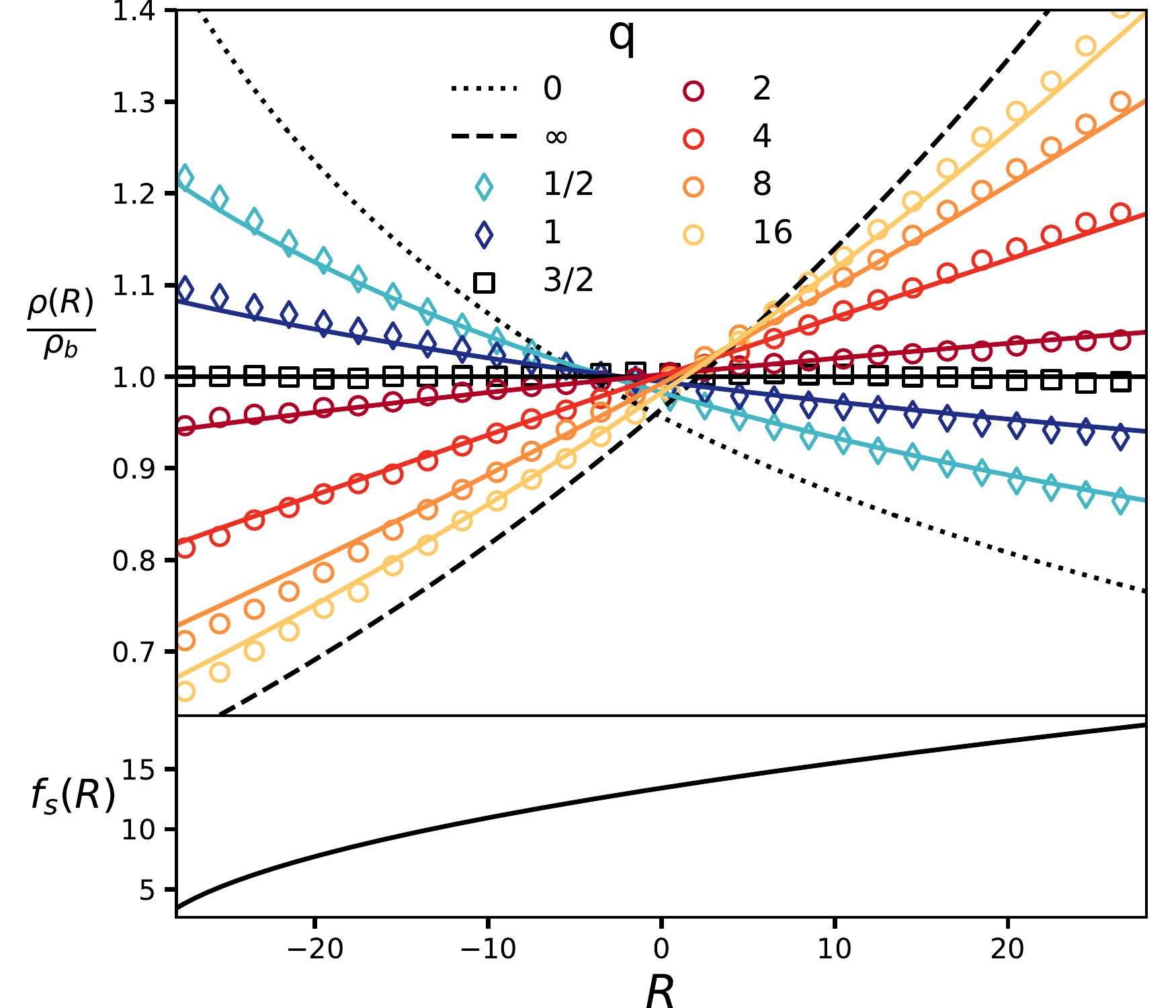} \caption{Density distributions of ABPs in three dimensions rigidly bound to
a passive cargo particle with different friction (top figure). Symbols
show simulation results; solid lines show the analytical prediction
(Eq. \eqref{dimer_density}). $\rho_{b}$ is the bulk density. The
swim force (activity) increases from left to right as $f_{s}(R)=\sqrt{6(R+30)}$
(bottom figure). With a highly mobile cargo particle (blue diamonds)
the dimer accumulates in regions of low activity (anti-chemotaxis).
At the predicted friction ratio $q=d/(d-1)=3/2$ a uniform concentration
is found (black squares). With a less mobile cargo (circles) the dimer
accumulates where the activity is high (chemotaxis). The dashed line
shows the limit of infinite friction of the cargo. This limiting case
corresponds to the behaviour of the system under the Born-Oppenheimer
approximation in which the cargo particle can be considered as immobile
and the ABP moving in a stationary potential. The dotted line shows
the limit of vanishing friction of the cargo, where the dimer behaves
like a single ABP. \label{fig:dumbbell_densities}}
\end{figure}

Figure \ref{fig:dumbbell_densities} displays the stationary density
distribution of dimers with a different friction ratio $q$. As long
as the cargo is highly mobile (small $q$), the dimer accumulates
in regions of low activity like a single ABP, which -- being an orthokinetic
swimmer -- is anti-chemotactic. When the cargo has large friction
(large q), it exhibits chemotaxis and accumulates where the activity
is high. The crossover from anti-chemotactic to chemotactic behaviour
can be quantitatively captured by our theory (Supplementary Information)
which yields the following expression for the steady-state density
of the dimer: 
\begin{align}
\rho(\RR)\propto\left[1+\frac{1}{1+q}\frac{\tau}{\gamma Td}f_{s}^{2}(\R)\right]^{-\frac{1}{2}\epsilon},\label{dimer_density}
\end{align}
where 
\begin{equation}
\epsilon=1-q\frac{d-1}{d}.
\end{equation}
For $q<d/(d-1)$, $\epsilon$ is positive, and the dimer is chemotactic.
For $q>d/(d-1)$, $\epsilon$ is negative, and the dimer is anti-chemotactic.
At the threshold $q=d/(d-1)$, the distribution of the dimer is uniform
and independent of the swim force. Note that in the limit of $q\rightarrow0$
the density distribution of the dimer reduces to that of a single
ABP, and it accumulates in the region of low activity. In general,
the exponent $\epsilon$ depends on the potential between the two
particles. In the Supplementary Information we show that the qualitative
behaviour of the system is unaffected by the choice of the potential.

In order to understand the mechanism behind the switch from anti-chemotaxis
to chemotaxis for increasing $q$, we consider the small and large
$q$ limits separately. When the friction coefficient of the cargo
particle is much smaller than that of the ABP, the cargo relaxes to
its quasi-steady state distribution at each position of the ABP. In
this limit, one can consider the dimer to be a single ABP with an
increased friction coefficient. Accordingly, the dimer accumulates
in the low activity regions (anti-chemotaxis). In the limit of large
friction of the cargo particle, the ABP relaxes to a quasi-steady
state distribution at each position of the load particle, and the
ABP has enough time to probe the neighbourhood of the cargo particle.
The activity gradient results in an effective force on the cargo particle
in the direction of the activity gradient. In this limit, one observes
the accumulation of the system in high activity regions (chemotaxis).

The steady-state density in the large friction limit can be obtained
following an independent approach similar to the Born-Oppenheimer
approximation in quantum mechanics. When the cargo is large, one can
consider it as immobile and the ABP as moving in a stationary potential.
The ABP explores the environment around the load and exerts an effective
force on it in the direction up the swim force gradient. This effective
force can be considered as the driving force for the total system.
The steady-state density of a passive Brownian particle with friction
$q\gamma$ in a such an effective force field is 
\begin{align}
\rho(\R)\propto\exp\left[\frac{1}{2}\frac{d-1}{d}\frac{\tau}{\gamma Td}f_{s}^{2}(\R)\right],
\end{align}
which shows that for large $q$ the dimer moves preferentially to
regions of high fuel concentration. For details of the calculations,
see the SI. Note that this density is the $q\rightarrow\infty$ limit
of equation~\eqref{dimer_density}. This consideration shows that
it is indeed the ability of the active particle to explore the activity
gradient that causes the chemotaxis.

\begin{figure}[t]
\includegraphics[width=1\columnwidth]{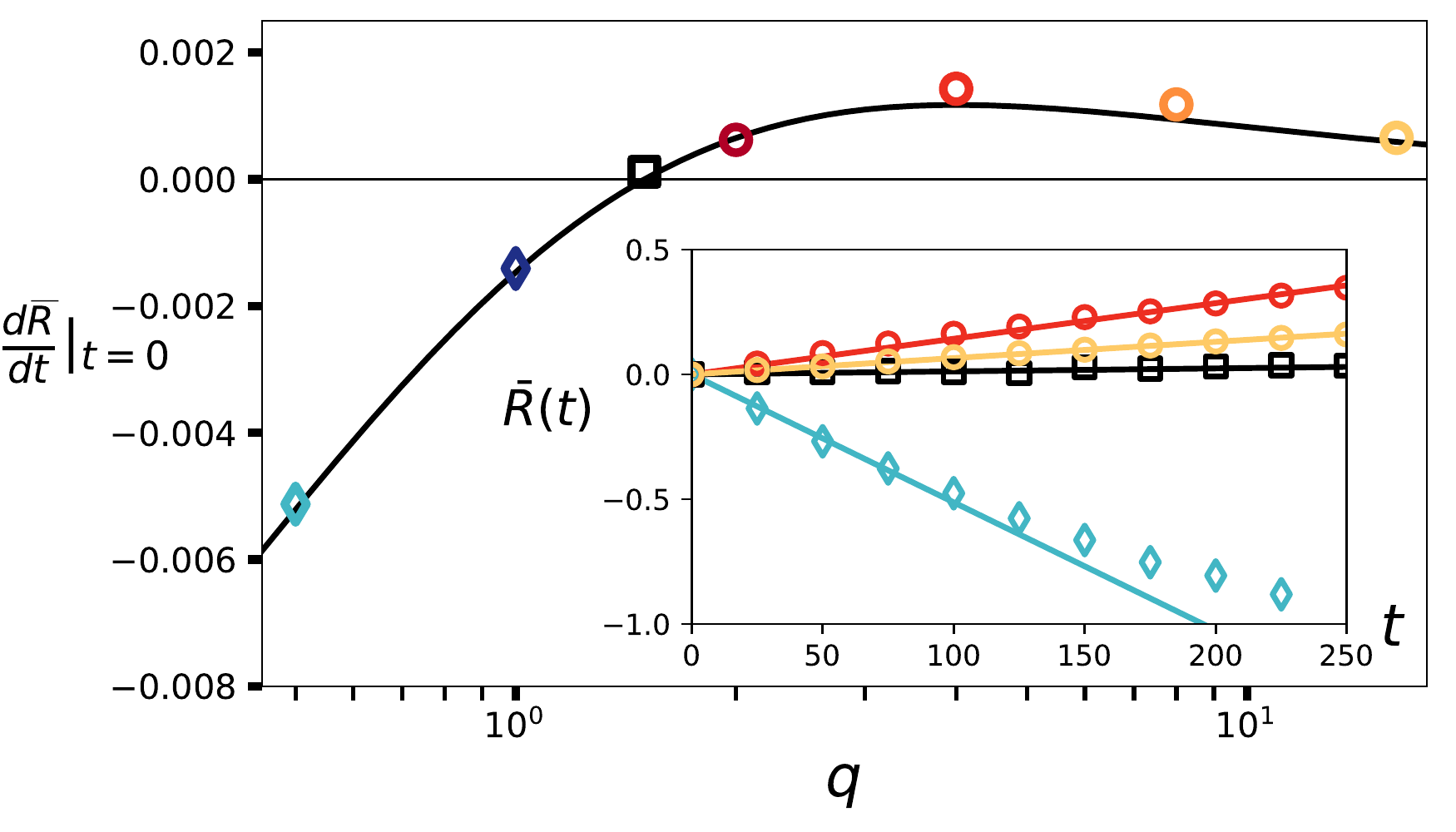} \caption{Time derivative at $t=0$ of the average position of the dimer starting
with a uniform distribution ($d\bar{R}/dt$) with swim force $f_{s}=\sqrt{6(R+30)}$.
Symbols show simulation results (colors as in Fig. \ref{fig:dumbbell_densities}).
The line is the theoretical prediction $d\bar{R}/dt\propto\epsilon/(1+q)^{2}$
with the proportionality constant fitted to the data (for details
see Supplementary Information). Inset: initial time evolution of the
average position ($R(t)$) for several values of $q$. The solid lines
show the linear fit for short time from which the data in the main
figure is extracted. For $q<3/2$ the dimers move to the left down
the swim force gradient. For $q>3/2$ the dimers move to the right
up the swim force gradient. As $q$ increases beyond $4$, the dynamics
start to slow down due to the increase in the friction of the dimer.
\label{fig:correlation}}
\end{figure}

Whereas the steady-state distribution measures the chemotactic behaviour
of the dimer, it does not shed any light on the `efficiency' of the
chemotactic transport of cargo particles by the ABPs. Though large
$q$ results in the strongest chemotactic behaviour, it also slows
down the transport of the cargo particle due to the increased friction
of the dimer. In our case, this leads to an optimum $q$ that yields
the fastest transport to regions of high activity. To quantify this,
we use the initial rate of change of the average position of the dimer,
starting with a uniform distribution (see Fig.~\ref{fig:correlation}).
At short times, the displacement of the dimer is determined by the
convective velocity (Supplementary Information), which is given by
$\boldsymbol{V}(\RR)\propto\nabla f_{s}^{2}\epsilon/(1+q)^{2}$. Depending
on the value of $q$, this is either up the swim force gradient (chemotaxis)
or down the swim force gradient (anti-chemotaxis). The convective
velocity has a maximum at $q=(3d-1)/(d-1)$ which coincides with the
simulation result. Note that for biased movement up the swim-force
gradient, only a large enough cargo is necessary, and no memory \citep{kromer_PRL20},
temporal integration of the fuel concentration \citep{cates_RProgPhys12},
or an explicit coupling between the swim-force gradient and the orientation
of the ABP \citep{lozano_NatureComm16,lozano2019diffusing} is required
.

\begin{figure}[t]
\includegraphics[width=1\columnwidth]{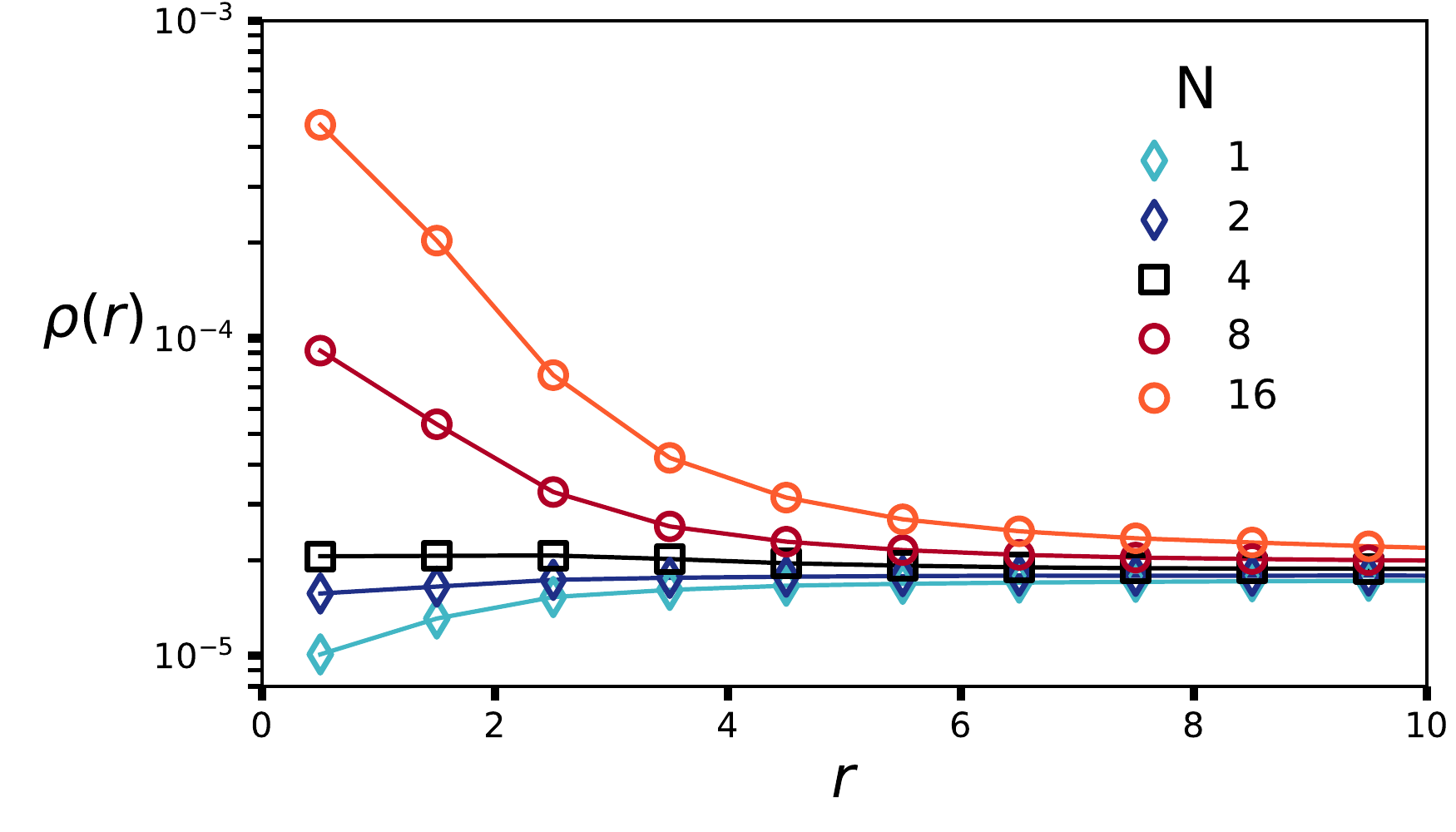} \caption{Stationary concentrations of ABP-molecules inside a spherical container
of radius $r_{max}=25$. The symbols show simulation data, and the
lines are a guide for the eye. The source of activity is at its center
and the driving force decreases as $f_{s}(r)=15/r$ for $r\ge1$,
and remains constant $f_{s}(r)=15$ for $r<1$. This activity profile
corresponds to a stationary fuel source emitting fuel by diffusion.
The mono- and dimers are anti-chemotactic and are driven out of the
region of high activity (blue diamonds) The quadromer is distributed
uniformly (black squares). Longer polymers ($N>4$, red circles) are
chemotactic and accumulate near the peak of the fuel concentration.
\label{fig:polymer_concentration}}
\end{figure}

Instead of coupling an ABP to a passive cargo, several ABPs can be
combined to form chains or clusters of $N$ 'monomers' ~\citep{kaiser2015does,winkler2020physics}.
Then, each individual ABP may be regarded as an active pulling agent
connected to a 'cargo' of $N-1$ particles because the other $N-1$
ABPs do not pull coherently in identical directions but rather in
random directions. Figure \ref{fig:polymer_concentration} displays
the stationary radial density distribution profiles of the molecules
inside the container. Short oligomers are anti-chemotactic and stay
away from regions of high activity. Long polymers, on the other hand,
are chemotactic and accumulate where activity is high. The crossover
point is seen with the quadromer which is roughly uniformly distributed
in the activity field.

Polymeric ABPs are therefore qualitatively similar to a single ABP
coupled to a passive cargo: If either the cargo is sufficiently inert
or the degree of polymerization sufficiently high, such that it inhibits
the free motion of an individual ABP, then the situation of a single
ABP inside an approximately stationary potential arises: Inside an
activity gradient, the particle is running up the confining potential
in the direction of increasing activity. The resulting positional
bias generates a net force which -- if the system is mobile -- drives
the same up the activity gradient. If, however, the cargo is rather
mobile, then the motion of the ABP remains approximately undisturbed
and the complex runs down the activity gradient as an individual ABP
does.

ABPs are equivalent to other models of active matter, such as run
and tumble ~\citep{cates_EPL13} active Ornstein-Uhlenbeck particles
~\citep{fodor_PRL16,martin2020statistical}, and the results shown
here also apply to these models. We show in the Supplementary Information
that the chemotactic behaviour observed in the active-passive dimer
cannot be reproduced in a system of dimers in which the active particle
is replaced by a passive Brownian particle coupled to a spatially
inhomogeneous thermostat.

An interesting outlook is the development of artificial nano-machines
which can locate origins of fuel gradients based on the design principle
of coupling ABPs to passive cargo without a complex sensing and steering
system. Furthermore, a natural progression of this work is to study
how the behaviour of a single active-passive dimer or more complex
clusters in an activity gradient affect the collective properties
of a system with a large number of such components \citep{liebchen2018synthetic,saha2014clusters,pohl2014dynamic,nasouri2020exact}.

The coupling of active bodies with or without passive bodies leads
to effective sensing of gradients of activity (fuel) and can be generalized
in several directions. First, the connectivity between the bodies
does not need to be linear as in the studies presented in this letter.
Virtually any kind of topology of connecting the active bodies leads
to effective chemotaxis, given that the complex provides sufficient
friction to single active bodies. Moreover, active bodies moving inside
a container that is permeable for the fuel but not for the active
bodies can give rise to chemotactic motion of the container, provided
the activity gradient is present inside the container. Also in this
case, it is the exploration of the active bodies in the gradient field
which leads to a higher averaged pressure on the container walls located
at the higher activity. Our results may shed light on the origin of
prebiotic forms of life and how chemotactic sensing has emerged during
evolution.

\bibliographystyle{unsrt}
\bibliography{active}

\section*{Acknowledgements}
A.S. acknowledges support
by the Deutsche Forschungsgemeinschaft (DFG) within the project SH 1275/3-1.

\section*{Author contributions}

All authors conducted the research together and participated in the
preparation and revision of the manuscript.

\section{Methods}

\label{methods}

\subsection{The Dimer Model }

Here we use the standard model of active Brownian particles ~\citep{bechinger_RevModPhys16}
coupled to a passive Brownian load particle:

\begin{eqnarray}
\partial_{t}\boldsymbol{r}_{1} & = & \frac{1}{\gamma}f_{s}(\rr_{1})\boldsymbol{p}+\frac{1}{\gamma}\F+\sqrt{\frac{2T}{\gamma}}\xxi_{1}\label{eq:ABP_Def_1}\\
\partial_{t}\boldsymbol{p} & = & \sqrt{2D_{r}}\boldsymbol{p}\times\eeta\label{eq:ABP_Def_2}\\
\partial_{t}\rr_{2} & = & -\frac{1}{q\gamma}\F+\sqrt{\frac{2T}{q\gamma}}\xxi_{2}\label{eq:ABP_Def_3}
\end{eqnarray}
where $\boldsymbol{r}_{1}$ denotes the position of the active particle
with friction constant $\gamma$ and swim speed $v(\rr_{1})$, $\boldsymbol{p}$
is a unit vector in the direction of the self-propulsion, and $\rr_{2}$
is the position of the passive load particle with friction constant
$q\gamma$ and $T$ is the temperature in units such that the Boltzmann
constant is one. The force binding the two particles together is $\F=-\nabla_{\rr_{1}}U(\rr_{1}-\rr_{2})$
and $D_{r}$ is the rotational diffusion coefficient. In this letter
we consider a stiff potential. The vectors $\eeta$, $\xxi$ and $\cchi$
are random Gaussian vectors with zeros mean and autocorrelation $\left<\xxi_{1}^{T}(t)\xxi_{1}(t')\right>=\left<\xxi_{2}^{T}(t)\xxi_{2}(t')\right>=\left<\eeta^{T}(t)\eeta(t')\right>=\boldsymbol{1}\delta(t-t')$.
The rigid bond between the particles is modeled as a harmonic potential,
\begin{equation}
U(\rr_{1}-\rr_{2})=\frac{1}{2}k\left(|\rr_{1}-\rr_{2}|-l_{0}\right)^{2}\;,\label{eq:bond-1}
\end{equation}
with equilibrium distance $l_{0}=1$ and spring constant $k$. To
make the bond rigid, we take the limit $k\rightarrow\infty$. In this
work we set $T=1$, the translational diffusion coefficient $D_{t}=1$,
$D_{r}=20$, and the frictional drag coefficient $\gamma=$1.

In this letter, we focus on the parameter range such that the persistence
lenght of the ABP is small compared to gradients in the of the swim
force $|\nabla f_{s}(\rr)|\tau/\gamma\ll1$, where $\tau=1/(d-1)D_{r}$
is the autocorrelation time of the orientation vector. Second, we
are interested in the parameter range where the separation between
the ABP and the load is small compared to gradients of swim force
$(\rr_{1}-\rr_{2})\cdot\nabla_{1}\ln f_{s}(\rr_{1})\ll1$. In this
regime the distribution of the ABP and that of the load are approximately
the same because the difference is related to the length scale of
the separation.

\subsection{Coarse graining the active-passive dimer}

The Fokker-Planck equation corresponding to Eqs. \eqref{eq:ABP_Def_1},\eqref{eq:ABP_Def_2}
and \eqref{eq:ABP_Def_3} is 
\begin{equation}
\partial_{t}P(t)=-\frac{1}{\gamma}\nabla_{1}\cdot\left[\F P(t)+f_{s}\pp P(t)-T\nabla_{1}P(t)\right]
\end{equation}
\[
+\frac{1}{q\gamma}\nabla_{2}\cdot\left[\F P(t)+T\nabla_{2}P(t)\right]+D_{r}\boldsymbol{\mathcal{R}}^{2}P(t),
\]
where $P(t)=P(\rr_{1},\pp,\rr_{2},t).$ We coarse grain this equation
in two steps. First we expand in eigenfunctions of the Laplacian on
the unit sphere $\boldsymbol{\mathcal{R}}^{2}$ \citep{cates_EPL13,vuijk2020lorentz},
and transform to the 'center-of-friction' coordinate $\RR=\frac{1}{1+q}\rr_{1}+\frac{q}{1+q}\rr_{2}$,
which we take as the position of the dimer, and the internal coordinate
$\rr'=\rr_{1}-\rr_{2}$. Then we integrate out the orientational
degrees of freedom ($\pp$) and the internal coordinate. This results
in the following equation for the probability density of the dimer:
\begin{equation}
\partial_{t}\rho(\RR,t)=-\nabla\cdot\boldsymbol{J},\label{eq:coarse_grained_FPE}
\end{equation}
where the flux of the dimers is
\begin{equation}
\boldsymbol{J}=-\frac{1}{2}\epsilon\rho(\RR,t)\nabla D(\RR)-D(\RR)\nabla\rho(\RR,t),
\end{equation}
and 
\begin{equation}
D(\RR)=\frac{1}{1+q}\frac{T}{\gamma}+\frac{1}{(1+q)^{2}}\frac{\tau}{d\gamma^{2}}f_{s}^{2}(\RR),
\end{equation}
is the course-grained space-dependent diffusion coefficient of the
dimer. The zero-flux steady-state solution to Eq. \eqref{eq:coarse_grained_FPE}
is Eq. \eqref{dimer_density} in the main text. Details of the calculations
can be found in the Supplementary Information.

\subsection{Brownian Dynamics Simulations}

For the simulations, a coarse-grained model is applied in which all
particles are ideal, i.e. no pair interactions except for bond forces
are implemented. In confined systems, particles interact with the
confining walls via a standard short-range repulsive Weeks-Chandler-Anderson
potential \citep{weeks_JCP71}. Bonds are harmonic potentials as in
Eq. \eqref{eq:bond-1}, with spring constant $k=170$ and equilibrium
distance $l_{0}=1$, which sets the unit length of the system. Due
to the large spring constant, this is essentially a bond with fixed
length.

The time integration of the stochastic differential equations is carried
out using a standard second-order Brownian dynamics algorithm~\citep{klenin_BiophysJ98}.
A tentative first-order displacement is
\begin{equation}
\rr_{i}'(t+dt)-\rr_{i}(t)=\frac{\F_{i}(\rr_{i},t)}{\gamma_{i}}\,dt+\sqrt{2D_{t}dt}\,\eeta_{i}(t)\;,\label{eq:first_step}
\end{equation}

with $i$ the particle index, $dt$ the time step and $D_{t}$ the
translational diffusion coefficient. The total force on the $i$th
particle, $\F_{i}$, is the sum of all conserved forces (from bonds
and bounding walls, if present) and, in case of ABPs, the driving
force 
\begin{equation}
\f_{i}(\rr_{i},t)=f_{s}(\rr_{i})\p_{i}(t)\;,\label{eq:f_drive}
\end{equation}
with the coordinates $\rr_{i}$, the coordinate-dependent activity
$f_{s}(\rr_{i})$, and the particle-embedded unit orientation vectors
$\p_{i}(t)$. The stochastic vectors $\eeta_{i}(t)$ are Gaussian
distributed with zero mean and time correlations $\langle\eeta_{i}(t)\,\eeta_{j}^{T}(t')\rangle=1\delta(t-t')\delta_{ij}$.
The second half-step is
\begin{equation}
\rr_{i}(t+dt)-\rr_{i}'(t+dt)=\frac{-\F_{i}(\rr_{i},t)+\F_{i}'(\rr_{i}',t+dt)}{2\gamma_{i}}\,dt\;,
\end{equation}
where $\F_{i}'(\rr_{i}',t+dt)$ are the forces calculated for the
advanced conformation $\rr_{i}'(t+dt)$. Finally, the orientation
vectors of the ABPs are updated via 
\begin{equation}
\p_{i}(t+dt)-\p_{i}(t)=\sqrt{2D_{r}dt\,}\eeta_{i}(t)\times\p_{i}(t)\;.\label{eq:rotation}
\end{equation}
For all simulations in this work, we used a timestep of $dt=10^{-3}$.

Because the theory does not incorporate boundary effects, we also
ignore these effects in Fig. \ref{fig:dumbbell_densities}. We do
this by removing the data closer than $\Delta$ to either of the walls
and rescaling the density such that $\int_{\Delta-L/2}^{-\Delta+L/2}dR\rho(R)=1$,
with $\Delta=$2.

\end{document}